\documentclass{elsart}

\usepackage{natbib}
\usepackage{longtable}
\usepackage{rotating}
\usepackage{soul}
\usepackage{array}
\usepackage{graphicx}

\usepackage{amssymb}
\usepackage{array}
\def\mnras{MNRAS}%
\def\aaps{A\&AS}%
\def\aap{A\&A}%

\def\apj{ApJ}
\def\nat{Nature}%
\def\jgr{J.~Geophys.~Res.}%
\def\apjl{ApJ}
\def\aj{AJ}
\begin{document}

\begin{frontmatter}



\title{Optical polarimetry of\\ Comet NEAT C/2001 Q4}


\author{S. Ganesh},
\author{U. C. Joshi},
\author{K. S. Baliyan\corauthref{cor1}}
\corauth[cor1]{Corresponding author}
\ead{baliyan@prl.res.in}

\address{Astronomy \& Astrophysics Division, Physical Research Laboratory,  Navrangpura, Ahmedabad 380009,
India}

\begin{abstract}

Comet NEAT C/2001 Q4 was observed for linear polarization using the optical
polarimeter  mounted at the 1.2m telescope at Mt. Abu Observatory, during the months
of May and June 2004. Observations  were  conducted through the International
Halley Watch narrow band (continuum) and $B,V,R$ broad band filters.  During the
observing run the phase angle ranged from $85.6^\circ$ in May to
$55^\circ$ in June. As expected, polarization increases with wavelength in this phase
angle range. Polarization colour in the narrow bands changes at different epochs,
perhaps related to cometary activity or molecular emission contamination.  
The polarization was also measured in the cometary coma at different locations along a line, in the
direction of  the tail.  As expected, we notice minor decrease in the polarization as
photocenter (nucleus) is traversed while brightness decreases sharply away from it.
Based on these polarization observations we infer that the comet NEAT C/2001 Q4 
has high polarization and a typical grain composition- mixture of silicates 
and organics.

\end{abstract}


\end{frontmatter}
\newpage
\section{Introduction}
\label{Intro}

Comets are considered to be primordial, icy bodies which are expected to retain
imprints of physical conditions prevailing in the proto-solar system. Therefore,
study of comets has proved very useful in our  understanding of the origin of the
solar system.   Sun light is scattered by the cometary grains and in this
process it gets polarized. The degree of polarization (P \%) mainly depends on: the
size distribution of the grains and their refractive index ($n - \iota k$);
wavelength of the light and the phase angle (sun-comet-earth angle, $\alpha$).
Linear polarization measurements have been made on many comets to
help understand their grain properties.  The first major efforts for detailed
polarization   observations were made for Comet P/Halley by many groups
\citep{Bastien1986,Brooke1987,Dollfus1987,Kikuchi1987,Lamy1987,
Leborgne1987,Metz1987,Sen1988}. There were other bright comets after P/Halley
such as  Comet C/1989 X1 (Austin), C/1996 B2 (Hyakutake), and C/1995 O1
(Hale-Bopp) for which polarization observations were reported (e.g.,
\citet{Sen1991,Joshi1997,Ganesh1998, Manset2000, Hadamcik2003}).
A database of comet polarimetry  has been created by \citet{Kiselev2005} which also
includes some unpublished data.\\

The study of the dust grains in comets has been an active area of investigation
for quite some time but the nature and composition of the cometary grains are
still not well understood. A study of the variation of polarization with phase
angle ($\alpha$) and the  wavelength dependence of
polarization helps to characterize the dust grains. Based on the data obtained
by various researchers, \citet{Dollfus1988}  established a typical
polarization-phase  curve describing the variation of the degree of linear
polarization  with the phase angle : small negative  values
of polarization for phase angle $\alpha$ $<$ 22$^\circ$, increasing nearly linearly in
the  range 30 $<$ $\alpha$ $<$ 70$^\circ$ \, and reaching a maximum value of
15-35 \% in the  phase angle range 90-110$^\circ$. \citet{Dollfus1989} pointed out  the
possibility of the grains giving rise to the polarization  being large, rough
and dark, resembling fluffy aggregates such as IDPs.  In-situ measurements 
\citep{Kissel1986,Mazets1986,LR1986,LR1999} by space
missions to Comet Halley  and corresponding numerical simulations by \citet{Fulle2000}
contributed important information on the nature of grains in that comet.
Return of the dust particles by
the Stardust Space Mission \citep{Brownlee2006}, has been a wonderful achievement in
the study of cometary material. The preliminary
studies suggest that most of the particles larger than a micron are
anhydrous (dry) silicates or sulfides. They include forsterite (Mg$_2$SiO$_4$), an
olivine end-member that is one of the first condensates from the solar nebula. 
Apart from their amorphous nature, some such silicate minerals are crystalline 
solids as also inferred from IR studies in comet Hale-Bopp \citep{Wooden1999} and Deep Impact mission comet 9P/Tempel 1
\citep{Lisse2006}.
The range of iron/magnesium ratios exhibited by the particles indicates that
comet Wild 2 is unequilibrated.  The comet contains an abundance
of silicate grains that are much larger than the predictions of interstellar grain
models, and many of these are high-temperature minerals that appear to
have formed in the inner regions of the solar nebula. Their presence in a comet
proves that the formation of the solar system included mixing on the grandest
scales \citep{Brownlee2006}. It is  remarkable that so many of the impacting comet particles contained at least 
a few relatively large solid grains, an order of magnitude larger than the size of 
typical interstellar grains \citep{Mathis1977}.
In addition to silicates and abundant sulfides, the collected comet samples
contain organic materials \citep{Sandford2006}.
The mixing of material from inner and outer solar nebula as indicated by the Wild 2 coma 
samples returned by Stardust and the great heterogeneity in comet nucleus make up shown 
by the gas and dust coming out of the Deep Impact comet 9P/Tempel 1
\citep{Schmitz2008, Ahearn2006, Harker2005}, make it more challenging to pin down the
region of the formation of comets. It also underlines the need for more studies in 
this direction.

Space experiments are very expensive and demand long lead time. Therefore, data collected 
using ground based facilities play a significant role in this study.  Ground based 
observations have indicated the detailed behavior of grains to change from one comet 
to another. This fact is corroborated by the Deep Impact, Startdust and comet Halley 
space missions also. At a basic level, the size distribution of impacted Wild 2 grains is 
different from the ejecta of the Tempel 1 \citep{Lisse2006} and even that seen in 
the comet Halley.  For example, as per Deep Impact results, Comet 9P/Tempel contains MgFe
sulfides and carbonates,  at odds with Stardust samples. These sulfides are also absent in IDPs. So,
either the comet compositions are different or sampling regions are different. Another
possibility could be improper laboratory constituents used to model Deep Impact spectra
\citep{Brownlee2006}.
Until more in-situ measurements are carried out on many other comets, the 
information about dust grains in the comets has to come largely from ground based 
observations in conjunction with theoretical models/simulations
\citep{Krisnaswamy1986,Xing1997,Jockersetal1997,Petrova2001b, Petrova2001a}.\\

On the basis of the polarization behavior of 13 comets,  \citet{Chernova1993}
pointed out the existence of two types of comets: gassy and dusty.
Subsequently \citet{LR1996} and \citet{Hadamcik1997} expanded this work by
studying the polarization phase curve for a large  number of comets. Quantitatively, for
$\alpha$ $>$ 30$^\circ$ \,  it was thought that comets follow two distinct
distributions in the polarization-phase angle diagram leading to two classes of comets:
low polarization and high polarization
comets \citep{LR1996} with high polarization comets being dusty and low
polarization comets gassy. \citet{Kolokolova2007} present a strong argument in favour of the two
classes of comets.  They claim a strong correlation of the polarization
behaviour with the strength of the 10$\mu$m silicate feature which is a direct
signature of the dust grains.  However, it is to be noted that cometary
classification by IR spectra and by polarization values is highly correlated \citep{Kolokolova2004}.
Any comet may belong to either class depending upon the dust activity. 
Also, the differences in the extent of polarization in 
the two classes could be due to molecular emission contamination
\citep{Jockers2005}.    In case of the gassy comets, the emission from molecules
covers large fraction  of the spectrum and  contaminates the continuum,
thus lowering the degree of polarization.   \\

The apparition of comet NEAT C/2001 Q4 provided a good opportunity to
make polarimetric observations at  large phase angles  which are useful to study
cometary composition. During the
observing run in May-June 2004, the comet was conveniently located in
the sky and was bright enough to achieve good S/N ratio. In
this communication we report the polarization observations made in
May 2004 with the phase angle range  $86-79^\circ$, and
later in June with phase angle range $ 55-56^\circ$ and discuss the  results
obtained.\\

\section{Observations and analysis}

Photopolarimetric observations of comet NEAT C/2001 Q4 were made during the
period May-June 2004  with a two channel photopolarimeter
\citep{Deshpande1985,Joshi1987} mounted on the 1.2m telescope of Mt. Abu
Observatory operated by Physical Research Laboratory (PRL), Ahmedabad. PRL
photopolarimeter modulates the polarized component of the incident light at
20Hz with a rotating  super-achromatic half-wave plate in front of a Wollaston
prism. The instrument is equipped with IAU's International Halley Watch
($IHW$) continuum filters (3650/80\AA, 4845/65\AA, 6840/90\AA~) \citep{Osborn1990}  and Johnson-Cousins' $BVR$ broad band filters. The
$IHW$ filters acquired for Comet Halley have been
in regular use for observations of several other comets
\citep{Joshi1987,Sen1991,Ganesh1998,Joshi2003}. The filters have been carefully
stored in dry atmospheric conditions to preserve their transmission
characteristics.  The observations made with the same set of filters facilitate
better comparison with other comets observed  earlier,  hence their continued
use. \\


\begin{sidewaystable}[h]
\caption{\label{obslogtab}Observation log and comet parameters; 4th column represent the observation time:
 beginning and end are given in first and second row along with corresponding RA, DEC, Heliocentric and Geocentric distances in subsequent columns.}


\begin{tabular}{ccccccccc}
\hline
Date &	Moon-rise& Sun-set  &UT &RA(2000) &Dec(2000) & r & $\Delta$&
$\phi$(SCE)\\
\hline

2004:05:13 & 21:02 & 13:47 & 15:08 & 8:23:16.17 & 13:43:25 & 0.96283301 & 0.38909076 & 85.639\\
	   &       &       & 15:22 & 8:23:20.00 & 13:45:31 & 0.96282576 & 0.38926712 & 85.632\\
2004:05:14 &       &       & 14:43 & 8:29:38.26 & 17:05:21 & 0.96225326 & 0.40752694 & 84.872\\
	   &       &       & 15:10 & 8:29:45.21 & 17:09:00 & 0.96224519 & 0.40789635 & 84.857\\
2004:05:15 &       &       & 14:50 & 8:35:44.56 & 20:12:30 & 0.96197946 & 0.42786269 & 83.996\\
	   &       &       & 15:21 & 8:35:52.02 & 20:16:18 & 0.96197713 & 0.42831640 & 83.977\\
2004:05:16 & 22:36 & 13:47 & 16:17 & 8:41:46.18 & 23:10:22 & 0.96204080 & 0.45070562 & 82.993\\
	   &       &       & 16:19 & 8:41:46.63 & 23:10:35 & 0.96204111 & 0.45073662 & 82.991\\
2004:05:17 & 23:08 & 13:47 & 15:30 & 8:46:56.12 & 25:36:42 & 0.96241040 & 0.47256353 & 82.020\\
	   &       &       & 16:14 & 8:47:05.47 & 25:41:05 & 0.96242694 & 0.47327385 & 81.989\\
2004:05:18 & \footnotemark{*}  & 13:47 & 15:10 & 8:51:53.52 & 27:51:45 & 0.96309469 & 0.49570033 & 80.987\\
	   &       &       & 15:48 & 8:52:01.10 & 27:55:10 & 0.96311726 & 0.49633426 & 80.959\\
2004:05:19 &new moon& 13:48& 15:12 & 8:56:37.81 & 29:55:49 & 0.96410640 & 0.51993316 & 79.905\\
	   &        &      & 16:04 & 8:56:47.58 & 30:00:03 & 0.96414883 & 0.52082508 & 79.866\\
2004:05:20 &15:00(set)&13:48& 15:50 & 9:01:12.08 & 31:50:51 & 0.96547329 & 0.54541115 & 78.771\\
	   &       &       & 16:25 & 9:01:18.29 & 31:53:25 & 0.96550969 & 0.54602547 & 78.744\\
2004:06:10 & 20:09 & 13:58 & 16:52 & 9:58:17.23 & 49:43:48 & 1.06098967 & 1.09889026 & 56.057\\
	   &       &       & 17:22 & 9:58:19.45 & 49:44:17 & 1.06114008 & 1.09942081 & 56.037\\
2004:06:11 & 20:40 & 13:58 & 16:55 & 10:0:03.92 & 50:07:05 & 1.06832835 & 1.12415131 & 55.120\\
	   &       &       & 17:02 & 10:0:04.43 & 50:07:11 & 1.06836446 & 1.12427407 & 55.116\\
\hline
\end{tabular}
\footnotetext{* observations were made untill the moon was up}
\end{sidewaystable}

\begin{figure*}[ht] 
\centering{
\includegraphics[width=\textwidth]{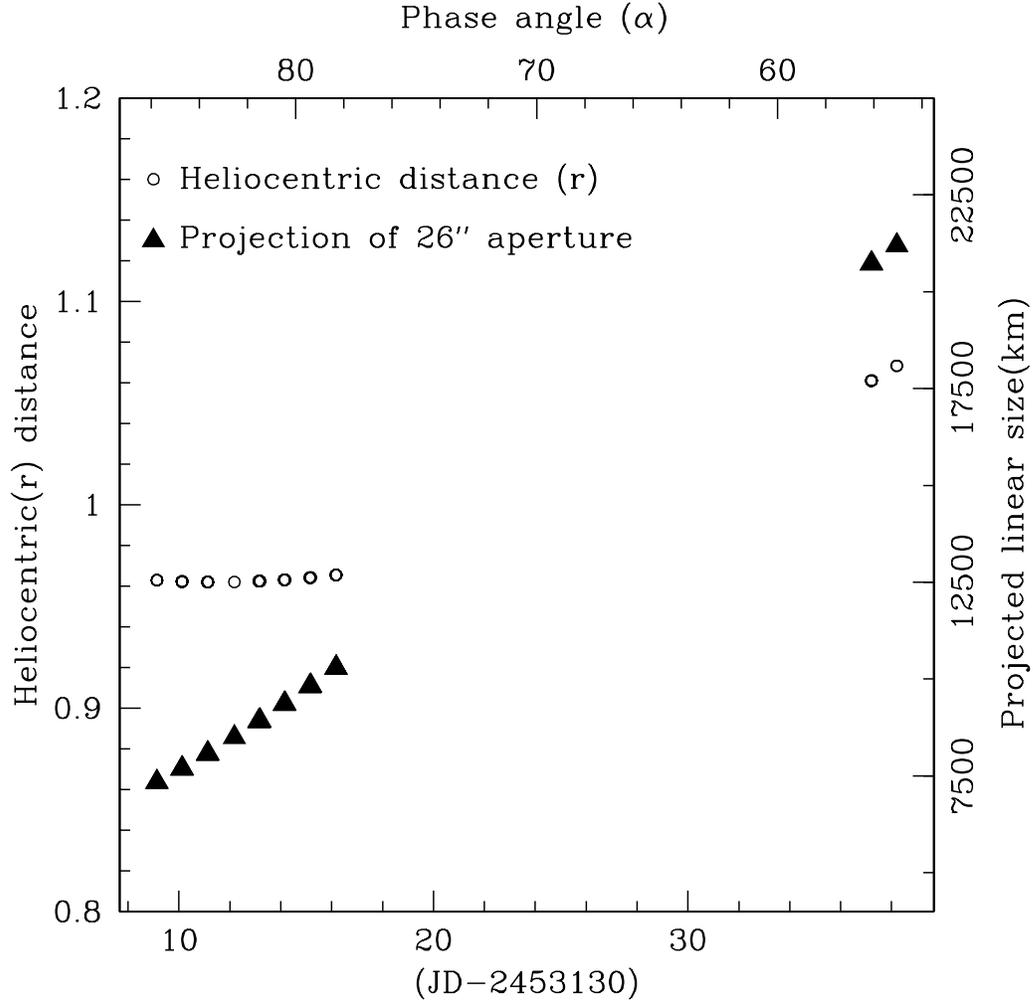}}
\caption{Apparent heliocentric range ($r$) (left axis) and linear size sampled at the distance of the comet (right axis) as a function of phase angle (top axis) and date of observation(bottom axis).}
\label{CometNEATHR}
\end{figure*}

Table \ref{obslogtab} shows the journal of the observations along with
other information like phase angle of comet, UT, moon rise/set time at Mt. Abu
etc. Heliocentric distance and the aperture projected on the
comet at the time of observations are plotted in Fig. \ref{CometNEATHR}. There
are eight nights of observations in May  and two in June 2004.
During May 13-16, the dark period of the sky being short, making
polarimetric observations on the comet in all the narrow band $IHW$ filters
with good S/N  was not possible. Therefore,
observations were made with broad band $BVR$ filters. Later on, when the
comet was available for longer dark-sky duration, 
the observations were made with $IHW$ filters. To facilitate
a comparison of the data taken with two sets of filters, observations on some
nights were made with both the sets of filters. After May 16, 
observations were made through two different apertures (cf Table \ref{obstab}) 
in each filter.  During the course of
our observations,  phase angle ranged between 55-86$^\circ$ 
(85.6-78.7$^\circ$ and 56.1-55.1$^\circ$ to be exact during May and
June respectively).  In Table \ref{obstab} mean values of the observed
parameters are listed.  Compared
to the background stars, comet was moving relatively fast (during the
observing run in May the estimated drift was $\sim$ 1" per 10 seconds)
which demanded frequent re-centering of the comet in the aperture. To
avoid significant drift of the comet within the aperture we kept
integration time short(typically 40sec) and also guided visually
through a finder telescope. Several such short exposures
were taken in each filter keeping the comet centered in the
aperture and the results were later averaged. In Table \ref{obstab}
mean values of polarization 
($P \% $), position angle($\theta$) and
error ($ {\epsilon}_P\% $)  are listed along with the integration time. A
least-squares fit to the data gives $P\%$ and $\theta$
while the error in the fit gives the error in $P\%$
 \citep{Joshi1987}. The errors in the position angle
are obtained using the equation 8.5.4 given by \citet{Serkowski1974}.

The observations were taken with apertures (non-metallic diaphragms centered on the photocentre
of the comet) of different sizes- 54''(a), 26''(b), 20''(c) and 10''(d).
All the comet observations were made during dark sky conditions, and the
comet being much brighter than the sky, sky polarization is negligible as compared to the comet degree of
polarization. Nonetheless, to take care of the sky polarization,
observations were made alternately centered on the photocentre of the comet and
towards a region of the sky more than 30 arcmin away from the comet
(along the anti-tail direction). It was noted visually that on nights 
subsequent to May 17, the comet became  more and more diffused 
 (the contrast between the nucleus, i.e. the photometric-center, and the
coma reduced with time). 
Sky conditions were photometric throughout the observations in May.\\

In comets, it is likely that the nature of dust (size and composition) changes as one moves
away from the nucleus.  To study this behaviour, observations were
made through different sized apertures. In order to explore this further, on May
20 comet was allowed to drift (starting from anti-tail side through nucleus to
the tail side) across an aperture of 10" diameter and data were sampled every
10sec. These observations were made using R-band filter.\\

Polarization standards 9 Gem, HR6353, HD183143, HD204827 were
observed to calibrate the observed position angle. Instrumental
polarization ($\sim$ 0.03\%) is much less compared to the
photon noise in the observations and hence is neglected.

\newpage
\setlength{\LTcapwidth}{\textwidth}
\setcounter{LTchunksize}{30}

{\scriptsize
\begin{longtable}{lcccccccccccc}
\caption[]{{\normalsize Polarization data for comet NEAT C/2001 Q4.  Listed
entries are,  date, time(UT), aperture (arcsec), Heliocentric ($r$) and Geocentric($\Delta$) distance(AU), projected diameter(kms), phase angle($\alpha$), filter (F),
total integration time (IT=integration time $\times$ number of integrations), degree of polarization($P\%$), error in polarization(${\epsilon}_P\%$),
position angle($\theta$ in degrees) in equatorial plane, brightness magnitude. \\
}}
\label{obstab} \\  
\hline
Date& UT &Ap&$r$&$\Delta$&Dia&  $\alpha$ & F &IT & $P\%$ & ${\epsilon}_p\%$ & $\theta$ & mag \\
\hline
\endfirsthead

\caption[]{Continued}\\
\hline
Date& UT&Ap&$r$&$\Delta$&Dia & $\alpha$ & F & IT & $P\%$ & ${\epsilon}_p\%$ & $\theta$ & mag\\
\hline
\endhead

\hline
\endfoot
13:5:04 &15:08:20 &  26" & 0.962833 & 0.389095 &7337& 85.6& B    &40*3 & 19.78 & 0.14 &18.5   &  6.40  \\
13:5:04 &15:16:00 &  26" & 0.962829 & 0.389191 &7338& 85.6& V    &20*3 & 21.39 & 0.46 &18.6   &  7.46  \\
13:5:04 &15:21:10 &  26" & 0.962826 & 0.389257 &7339& 85.6& R    &20*1 & 24.02 & 0.22 &18.2   &  5.11  \\
14:5:04 &14:43:40 &  26" & 0.962253 & 0.407536 &7684& 84.8& B    &30*2 & 19.68 &  0.18 &7.4   &  9.16  \\
14:5:04 &14:48:10 &  26" & 0.962252 & 0.407597 &7685& 84.8& V    &30*2 & 20.06 &  0.11 &8.7   &  7.58  \\
14:5:04 &14:52:20 &  26" & 0.962250 & 0.407654 &7686& 84.8& R    &30*2 & 23.92 &  0.10 &9.3   &  7.04  \\
14:5:04 &15:00:30 &  54" & 0.962248 & 0.407766 & 15970 & 84.8& B    &20*3 & 17.93 &  0.10 &9.7   &  8.19  \\
14:5:04 &15:05:00 &  54" & 0.962247 & 0.407828 & 15972 & 84.8& V    &20*2 & 18.34 &  0.05 &9.9   &  6.88  \\
14:5:04 &15:09:10 &  54" & 0.962245 & 0.407885 & 15974 & 84.8& R    &20*2 & 23.24 &  0.07 &9.9   &  6.55  \\
15:5:04 &14:51:10 &  26" & 0.961979 & 0.427880 &8068& 84.0& B    &20*2 & 18.57 &  0.11 &7.5   &  8.95 \\
15:5:04 &14:54:55 &  26" & 0.961979 & 0.427934 &8069& 84.0& V    &20*2 & 19.45 &  0.13 &7.7   &  7.80  \\
15:5:04 &14:58:15 &  26" & 0.961979 & 0.427983 &8070& 84.0& R    &20*2 & 23.07 &  0.11 &11.5  &  7.35  \\
15:5:04 &15:02:35 &  20" & 0.961978 & 0.428047 &6208& 84.0& B    &20*2 & 19.32 &  0.15 &7.5   &  9.45 \\
15:5:04 &15:05:50 &  20" & 0.961978 & 0.428090 &6209& 84.0& V    &20*2 & 20.07 &  0.14 &7.4   &  9.45  \\
15:5:04 &15:09:40 &  20" & 0.961978 & 0.428150 &6210& 84.0& R    &20*2 & 23.42 &  0.15 &15.1  &  7.85 \\
15:5:04 &15:13:40 &  54" & 0.961978 & 0.428209 & 16770 & 84.0& B    &20*2 & 17.34 &  0.10 &15.4  &  8.13  \\
15:5:04 &15:17:15 &  54" & 0.961977 & 0.428261 & 16772 & 84.0& V    &20*2 & 17.39 &  0.06 &11.1  &  6.98  \\
15:5:04 &15:20:50 &  54" & 0.961977 & 0.428314 & 16774 & 84.0& R    &20*2 & 22.54 &  0.13 &7.9   &  6.68  \\
16:5:04 &16:18:30 &  26" & 0.962041 & 0.450729 &8498& 83.0& 3650 &40*6 & 17.76 & 2.08  &4.8   & 9.38  \\
17:5:04 &15:30:44 &  26" & 0.962411 & 0.472575 &8911& 82.0& 3650 &40*7 & 15.32 & 1.20  &20.5  & 9.53  \\
17:5:04 &15:37:25 &  26" & 0.962413 & 0.472683 &8912& 82.0& 4845 &40*6 & 20.80 & 0.34  &14.1  & 8.28  \\
17:5:04 &15:43:40 &  26" & 0.962415 & 0.472784 &8914& 82.0& 6840 &40*5 & 25.36 & 0.33  &14.0  & 7.19  \\
17:5:04 &16:14:30 &  20" & 0.962427 & 0.473282 &6864& 82.0& 4845 &40*6 & 21.21 & 0.43  &13.9  & 8.91  \\
17:5:04 &16:14:50 &  20" & 0.962427 & 0.473287 &6864& 82.0& 3650 &40*7 & 17.53 & 2.80  &13.7  & 10.14  \\
17:5:04 &16:15:24 &  20" & 0.962427 & 0.473295 &6864& 82.0& 6840 &40*9 & 25.28 & 0.52  &13.9  & 7.85  \\
18:5:04 &15:10:20 &  26" & 0.963095 & 0.495706 &9347& 81.0& 3650 &40*5 & 19.31 & 1.84  &4.6   & 9.42  \\
18:5:04 &15:22:30 &  26" & 0.963102 & 0.495909 &9350& 81.0& 4845 &40*4 & 21.01 & 0.17  &5.9   & 8.15  \\
18:5:04 &15:30:50 &  26" & 0.963107 & 0.496048 &9353& 81.0& 6840 &40*6 & 24.76 & 0.47  &6.2   & 7.21  \\
18:5:04 &16:05:50 &  20" & 0.963128 & 0.496632 &7203& 81.0& 3650 &40*7 & 17.03 & 2.55  &13.4  & 10.06  \\
18:5:04 &15:57:15 &  20" & 0.963123 & 0.496489 &7200& 81.0& 4845 &40*6 & 21.72 & 0.50  &6 2   & 8.83  \\
18:5:04 &15:47:20 &  20" & 0.963117 & 0.496323 &7199& 81.0& 6840 &40*7 & 24.54 & 1.22  &5.3   & 8.02  \\
19:5:04 &15:14:00 &  26" & 0.964108 & 0.519967 &9804& 79.9& 3650 &40*7 & 17.81 & 2.55  &1.6   & 10.18  \\
19:5:04 &15:35:30 &  26" & 0.964125 & 0.520336 &9811& 79.9& 4845 &40*5 & 20.88 & 0.49  &7.8   & 8.96  \\
19:5:04 &15:47:30 &  26" & 0.964135 & 0.520542 &9815& 79.9& 6840 &40*9 & 23.98 & 0.49  &7.4   & 7.99  \\
19:5:04 &16:20:00 &  20" & 0.964162 & 0.521100 &7558& 79.9& 3650 &40*4 & 18.15 & 4.04  &27.7  & 10.92  \\
19:5:04 &16:11:45 &  20" & 0.964155 & 0.520958 &7556& 79.9& 4845 &40*5 & 20.84 & 0.72  &5.5   & 9.73  \\
19:5:04 &16:02:45 &  20" & 0.964148 & 0.520804 &7554& 79.9& 6840 &40*4 & 24.06 & 0.85  &7.7   & 8.62  \\
20:5:04 &15:52:10 &  20" & 0.965475 & 0.545449 &7911& 78.8& 3650 &40*5 &  20.22 & 3.05 &9.3   & 10.34  \\
20:5:04 &16:00:20 &  20" & 0.965484 & 0.545592 &7913& 78.8& 4845 &40*5 &  19.38 & 0.51 &7.9   & 9.22  \\
20:5:04 &16:08:35 &  20" & 0.965492 & 0.545737 &7915& 78.8& 6840 &40*5 &  23.27 & 0.64 &7.5   & 8.17  \\
20:5:04 &16:14:50 &  20" & 0.965499 & 0.545847 &7917& 78.8& B    &10*5 &  16.70 & 0.25 &7.7   & 9.76   \\
20:5:04 &16:20:45 &  20" & 0.965505 & 0.545951 &7919& 78.7& V    &10*5 &  16.59 & 0.09 &6.8   & 8.79   \\
20:5:04 &16:24:50 &  20" & 0.965509 & 0.546022 &7920& 78.7& R    &10*5 &  20.50 & 0.17 &7.8   &  8.39  \\
10:6:04 &16:54:40 &  26" & 1.061004 & 1.098940 &20722& 56.0& V    &60*4 &   9.92 & 0.13 &9.5   & 9.78   \\
10:6:04 &16:43:15 &  26" & 1.060944 & 1.098735 &20718& 56.1& R    &60*3 &  12.50 & 0.25 &8.8   & 9.38  \\
10:6:04 &17:05:30 &  26" & 1.061057 & 1.099129 &20726& 56.0& 4845 &60*7 &  14.29 & 1.08 &5.8   & 10.65  \\
10:6:04 &17:20:50 &  26" & 1.061134 & 1.099400 &20731& 56.0& 6840 &60*10&  16.04 & 1.71 &9.4   & 9.48 \\
11:6:04 &16:56:10 &  26" & 1.068334 & 1.124172 &21198& 55.1& B    &20*3 &   7.64 & 0.53 &26	 &  11.94\\
11:6:04\footnote{sky condition : scattered cloud patches} &17:01:40 &  26" & 1.068363 & 1.124268 & 21200 & 55.1& V    &20*2 &  14.50 & 0.99 &43    & 11.37\\

\end{longtable}
}
\newpage

\section{Results \& discussion}

Results from continuous photopolarimetric observations during May 13-20, and
June 10-11, 2004, covering the comet in pre and post perihelion
passages, are presented in Table \ref{obstab} in the form of polarization percentage
($P\%$), error in $P~ ({\epsilon}_p\%$), position angle ($\theta$) and magnitude
at different phase angles, filters and aperture sizes.

In order to study the radial distribution of polarization, and
hence the dust particles, we made observations through different sized
 apertures. In the case of Comet Halley, a fair degree of
agreement is seen among the observations made by different groups with
different size apertures as long as the polarization is estimated
with apertures large enough to cover good part of the coma,
centered on the nucleus, which averages out the effect of heterogeneity.  
In the case of comet Hale-Bopp also no significant difference  was found 
in the polarization observed through two different  apertures corresponding
to linear scales of 14318km and 28313km respectively \citep{Ganesh1998} and 
similar results were reported by \citet{Manset2000}.

Figure 1 shows variation of the Heliocentric range and linear size sampled 
at the geocentric distance of the comet (corresponding to 26" aperture)
during the observing run in May and June. The Heliocentric range ($r$) of
the comet has not changed much during the observing run in May while it
changes slightly in Jun 10-11, 2004. The total change in $r$ during the
observing run (May-June) is less than 10 \%. However, most of the observations
are made in May and the change in $r$ during this period is less than 0.4\%,
thus resulting in a very small change in solar flux as seen by the comet. On
the other hand the Geocentric distance ($\Delta$) has changed rapidly during
the observing run, resulting in large change in the sampled area projected on
the comet surface. Our aperture of 26" corresponds to a projected diameter
ranging from $\sim$ 7350 to $\sim$ 21800kms over the course of the observations in
May and June. Therefore the effects of  inhomogeneities in the coma are expected to be
averaged out. The observed data in Table 2 support this view. We also note that
the overall polarization characteristics of NEAT C/2001 Q4 are similar to comet
Halley and C/2000 WM1 (LINEAR) as is discussed later. Therefore, the
comparison of the polarimetric observations on different dates for NEAT C/2001
Q4 is meaningful.

\subsection{Wavelength dependence of polarization}

\begin{figure*}[ht] 
\centerline{\includegraphics[width=\textwidth]{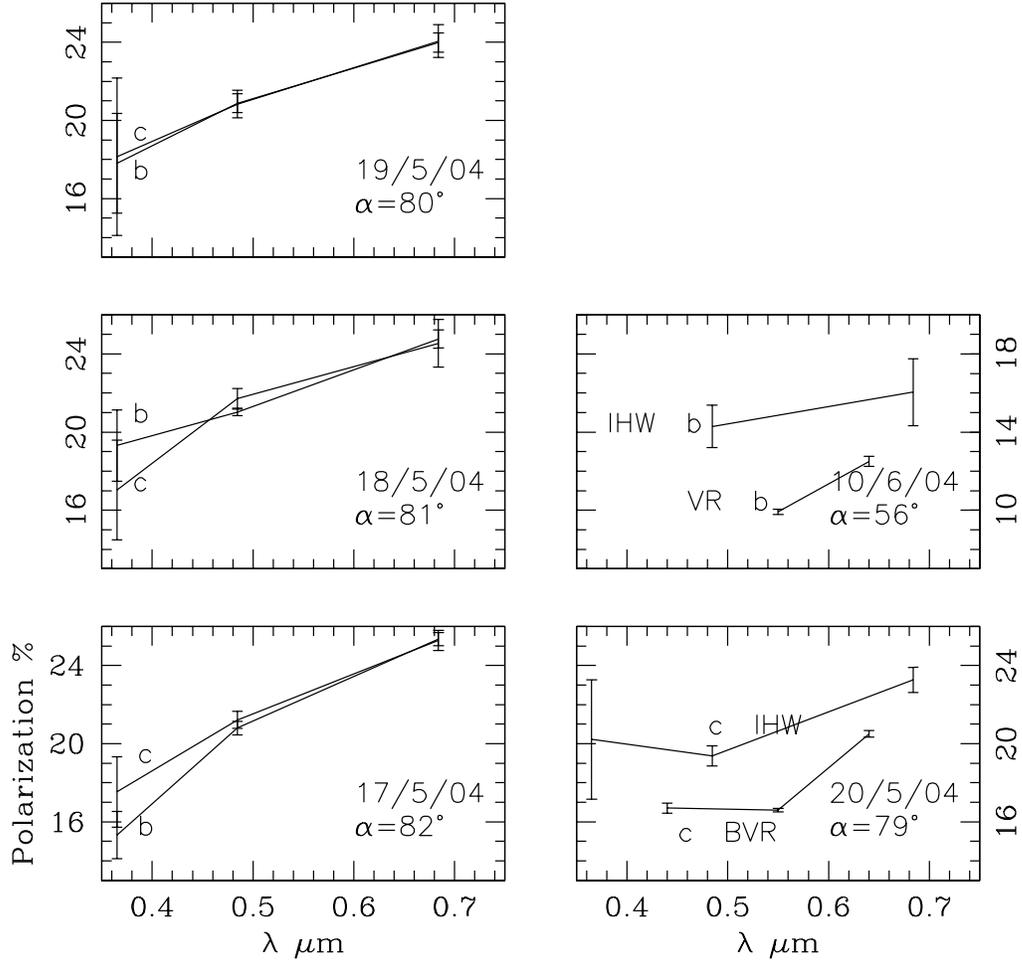}}

\caption {Wavelength ($IHW$ narrow band) dependence of the degree of polarization for
the comet NEAT.  Polarization observed through different apertures is marked with letters 'b', 'c' representing
the apertures 26'' and 20'' respectively.  At phases $\alpha=56$ and $\alpha=79$ (panels in right column) observations are plotted in both broad and narrow bands (see text).} \label{wdep}

\end{figure*}

Wavelength dependence of polarization as observed in $IHW$ filters on various
dates in May and June is plotted in Fig. \ref{wdep}.  Observation date, phase 
angle and the apertures used are marked in the figure. As mentioned in earlier section, apertures $a, b, c$ stand for 54'', 26'', 20'', respectively. The figure shows
that the degree of polarization increases with the wavelength. 
The errors are relatively large for
3650\AA~ $IHW$ filter, but the trend of increase in polarization with
wavelength is clear.  Similar behaviour is noted by \citet{Kiselev1998} for comet
Hyakutake and by \citet{Manset2000} for comets Hale-Bopp and Hyakutake who also report
polarization observations at large phase angles.

\begin{figure*}[ht] 
\centerline{\includegraphics[width=\textwidth]{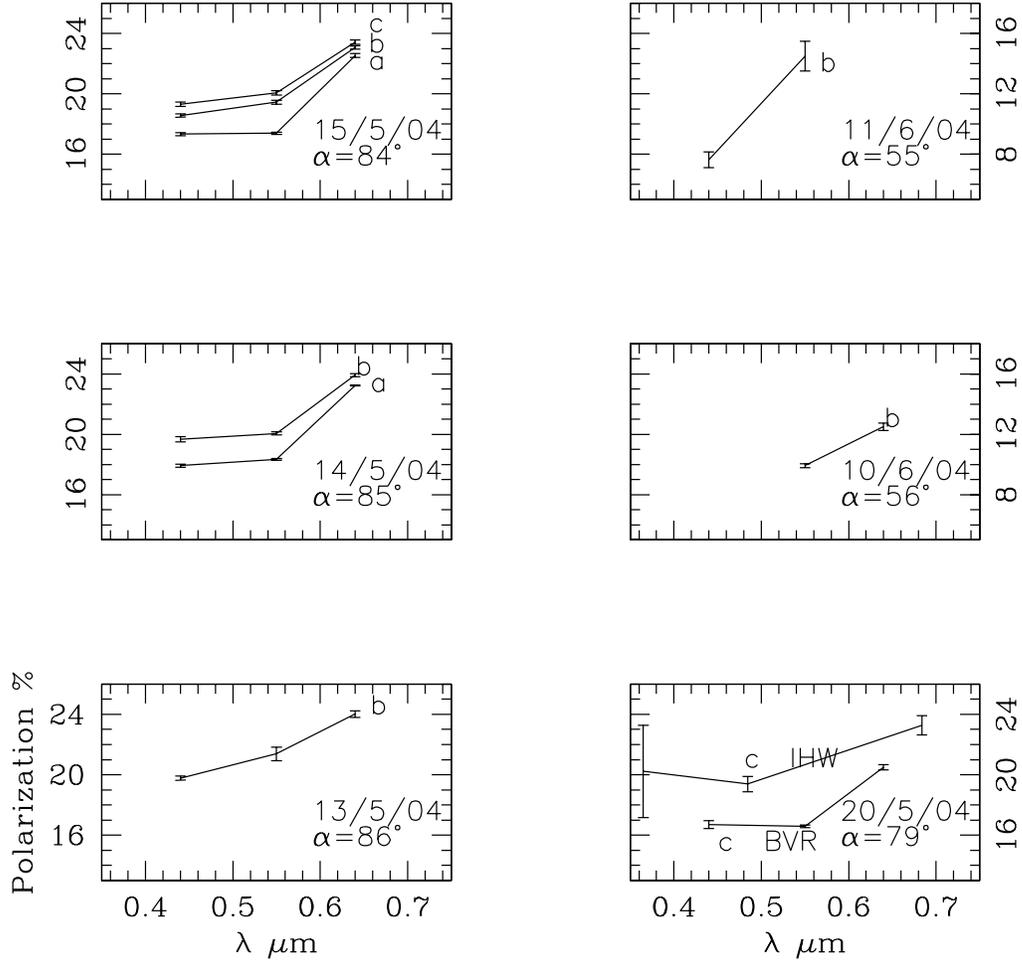}}
\caption {Wavelength dependence of the degree of polarization for the  comet NEAT in the $BVR$ broad bands.
The  aperture 'a' is of size 54'' and the apertures 'b', 'c' are the same as in Fig. \ref{wdep}.}
\label{wbvrdep}
\end{figure*}

On May 20, observations have been made through both the sets of filters ie $IHW$ and
$BVR$ and plotted in the panel labeled 20/5/04 of Fig. \ref{wdep}.  
The values of polarization for broad band filters are lower compared to the $IHW$
filters. This is expected as the broad bands usually encompass several emission
lines. The $R$-band is least contaminated by molecular lines while $V$-band contains
the C$_2$ emission band and the $B$-band is contaminated by CN-band emission.
Polarimetric observations on May 13-16, 20 and June 10-11 are made through $BVR$
filters. It is clear from the  Fig. \ref{wbvrdep} that there is a noticeable change in 
the polarization in $V$-band 
at different dates while the other two bands ($BR$) do not show such pattern.  
This indicates significant variation in the molecular emission (eg C$_2$ band emission) in $V$ band.

Considering the behaviour of polarization in apertures of different sizes, in
general, the degree of
polarization and its wavelength dependence do not depend significantly on
the sampled size of the comet  unless size difference is really large. For example,
\citet{Manset2000} report a decrease of 2\% in polarization in the coma of comet Hale-Bopp
in going from 8'' to 80'' aperture size. At the same time, they do not
find any change in the polarization values obtained through  8.2'' and 10.6'' 
apertures. As explained in the earlier
section, we also do not expect any appreciable change in the 
polarization behaviour in the apertures used. On May 17 polarization is marginally
lower at shorter wavelengths {\it in larger aperture} while on May 18, 
there is a relative increase in the polarization in 4845\AA~band and a decrease in 
$3650\AA$~band in smaller aperture. However, the errors in polarization are large in the later
case (cf Figure 2) and  therefore any definite conclusion might be misleading here. 
Nonetheless, if the change in polarization behaviour in two apertures is
genuine, it could be due to cometary activity, perhaps before May 17 resulting in temporary
non uniform distribution of dust. Subsequently, the grain distribution gets
homogenized after the activity has subsided, leading to similar wavelength
dependence of polarization seen through different apertures on May 19.  This
also corroborates well with our visual observations of slightly diffused
appearance of the comet on May 19 and 20  leading to slightly lesser contrast between
the photocentre and background coma. However, it is worthwhile to note that 
non-variation of polarization with aperture does not rule out presence  
of any other activity in comet. 

\begin{figure*}[ht] 
\centerline{\includegraphics[width=\textwidth]{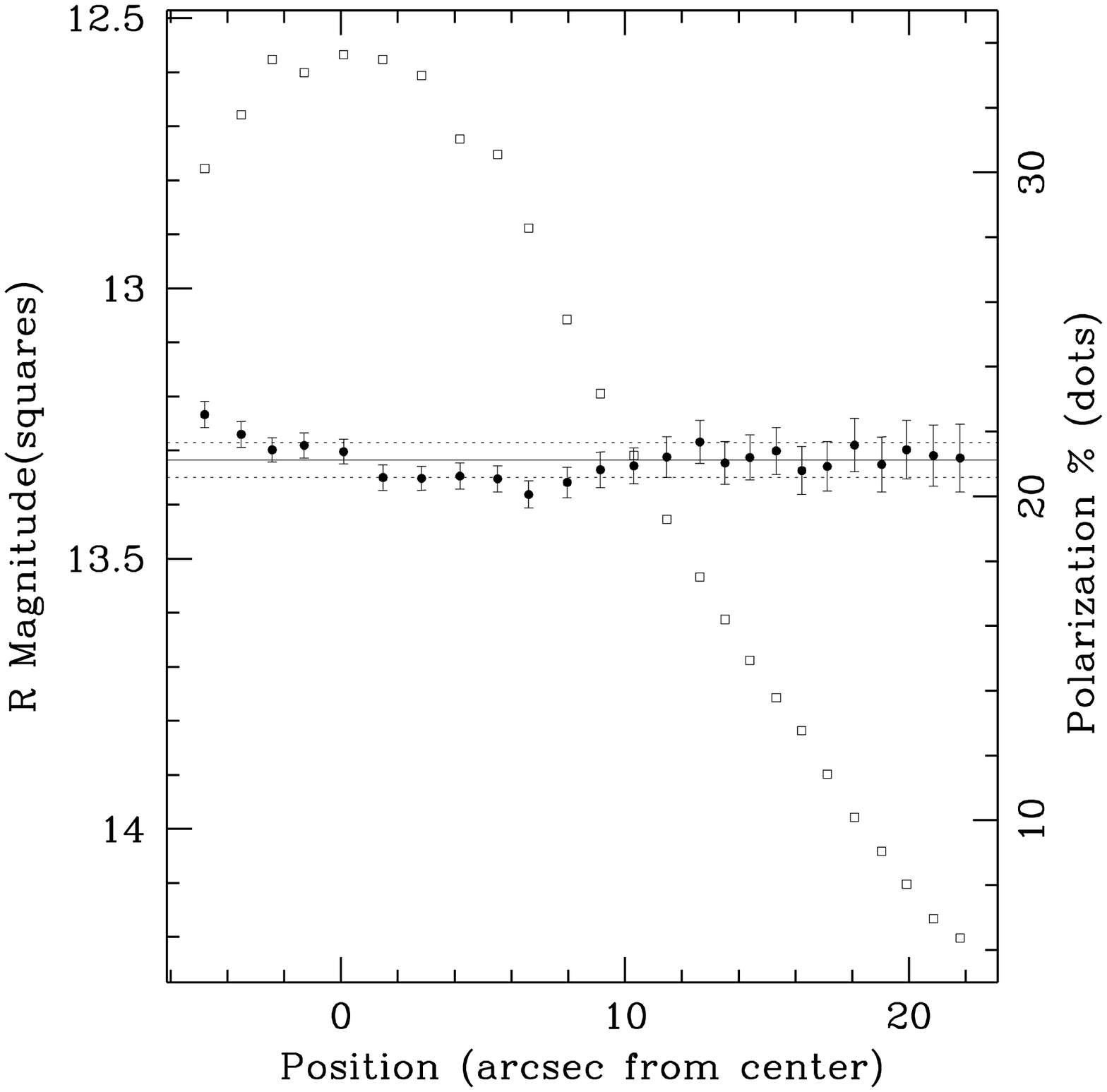}}
\caption{Variation of polarization and instrumental magnitude with location in a 10'' aperture. See text for details}
\label{drift}
\end{figure*}

Now let us discuss how the polarization colour (spectral gradient of polarization) 
changes with time. We computed polarization color in $IHW$ continuum bands using the
expression:
 \begin{equation} PC =  {(P_r - P_b) \over
(\lambda_r - \lambda_b)} 
\end{equation}

 Where $P_r \ and \ P_b$ are polarization values at $0.6840\mu $m and $0.4845\mu $m, 
respectively. The polarization colour (PC) and photometric colours $ \delta_{Mag}(b - r) $ 
in two apertures are listed in Table 3 on different dates.  
We note that PC  decreases during May 17 to May 19 in 26'' aperture while in smaller
(20'') aperture, it decreases on 18th May but increases during 19 to 20th May 2004. The 
photometric colour also decreases on 18th May and then shows an increase in both the
apertures. That means that coma of comet NEAT Q4 gets bluer in polarization as well as
in photometric colour during 17-18 May 2004. An increase in blue colour in intensity together
with increase in linear polarization is generally interpreted as decrease in the average
grain size distribution.  It could mean an 
increase in the number of fluffy (porous) aggregate particles and/or finer grains due to
sublimation of organics or some other activity as referred to earlier also. 
It could also indicate a change in composition due to decomposition of organics 
since at about 1 AU heliocentric distance, it is the  dominant process in near-nuclear region \citep{Kolokolova2001}.
 On May 19, polarization and photometric colors have 
increased again in small aperture, indicating,  probably, a fresh activity on the comet. 
Such correlation can be explained by sublimation of some absorbing material- such as dark organics. 
However, present observations do not allow us to make any definitive statement 
about the type of activity.  

Considering the polarization values in $IHW$, we also note that the wavelength
dependence of polarization is steeper at larger phase angle(82$^\circ$)
compared to what it is at lower phase angle(55$^\circ$) ie polarization colour
is redder at higher phase angle.

\begin{table}[h]
\caption{Polarization colour(PC) and photometric colour \newline
 ([4845]-[6840]) in 26 \& 20 '' apertures}

\begin{tabular}{l|cc|cc}
\hline

Date &\multicolumn{2}{c|}{PC}&\multicolumn{2}{l}{Phot. colour}\\
\cline{2-3}  \cline{4-5}
&26''&20''&26''&20''\\
\hline\hline
17:5:04 & 22.86($\pm$ 0.47) & 20.40($\pm$ 0.67) &   1.09 &  1.06\\
18:5:04 & 18.80($\pm$ 0.50) & 14.10($\pm$ 1.32) &  0.94 &   0.81\\
19:5:04 & 15.54($\pm$ 0.69) & 16.10($\pm$ 1.11) &  0.97 &   1.11 \\
20:5:04 & ---   & 19.50($\pm$ 0.82) &   ---  &  1.05 \\
10:6:04 & 8.80($\pm$ 2.02)  &  --- &   1.17   &  --- \\
\hline

\end{tabular}
\label{pol_color}
\end{table}

\subsection{Spatial variation of polarization}
On 20th May 2004, we made polarimetric observations on the comet at various
points tail-ward. This was done by pointing the telescope ahead of the path of
the comet while tracking at the sidereal rate. Over a period of 360 seconds the
comet was allowed to drift across the 10'' aperture with sampling time of 10sec.  
During this period, the
movement of the comet was in a direction parallel to the direction pointed by
the tail. The observations were made using the R-band filter and  a running average of
5 measurements of polarization and intensity  are plotted in Fig. 4. 
An aperture of 10" corresponds to a
linear scale of $\sim$ 4000km at the distance of the comet at that time.
Any heterogeneity at scales larger than this should show up in the observed
polarization. We notice a decrease in polarization as the nucleus is approaching, becoming
near minimum at the photocenter while showing a systematic, albeit small, increase tail-ward.
Similar behaviour was noticed by \citet{Jewitt2004} for comet 2P/Encke, suggesting a change in the mean
scattering properties of dust particles on time of flight timescale of one hour- the
likely cause of fragmentation being disaggregation of the composite, porous dust particles. 
\citet{Manset2000} measured polarization in the coma and tail of comet Hale-Bopp
up to a distance of about 200'' and found the polarization to increase and then decrease
when going away from the nuclear region. In our case, perhaps the duration of observation and
linear distance covered were not sufficient enough to notice such variation. The
brightness reaches maximum, as expected, at the photocenter (nucleus) and then decreases
sharply. One can infer, broadly, the presence of large compact particles (aggregates) 
near the photocenter while more finer grains and fluffy aggregates tail-ward. 
However, too much should not be read in this behaviour as the sampled region
(4000 km) is larger than the typical size of the Comet nucleus and we might not have
sampled the circumnuclear region of lower polarization. 
The position angle does not show any appreciable change during the observation. 

\subsection{Polarization phase curve}
The variation of polarization as a function of phase angle throws light on various properties
of comet such as composition, albedo, size distribution of grains etc. Though there exists
large amount of observational data related to polarization at many phase angles, polarization
values at lower phases (negative polarization branch region) and larger phase angles,
including the region where it peaks (90 - 105$^o$) are very scanty. Such scarcity of data
causes problems for the proper modeling/fitting of the polarization phase curve for the
comets.

\begin{figure*}[ht]
\centering{
\includegraphics[width=\textwidth]{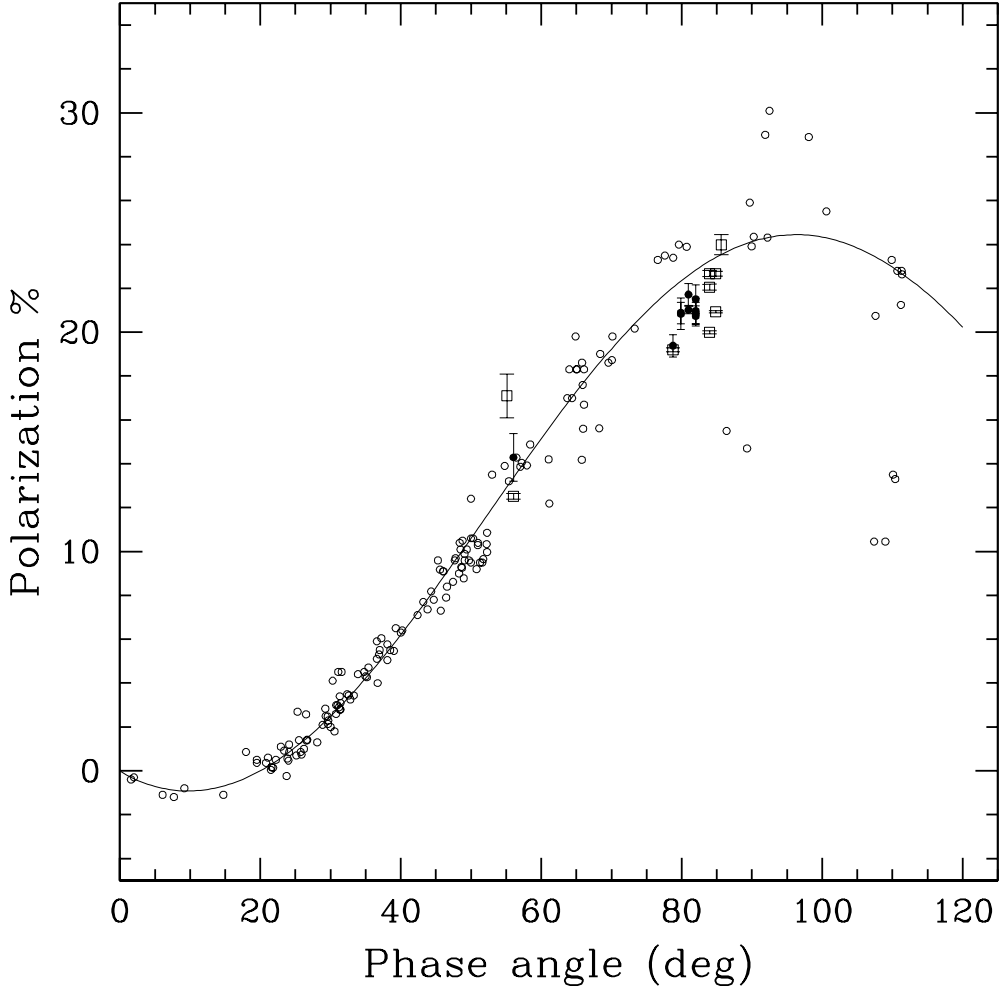}}
\caption{The phase dependence of polarization of Comet NEAT C/2001 Q4
for blue wave bands centered at 4845\AA~ are shown as black solid dots with error bars.  
Observations in $V$ broad band are shown as open squares (with an additional offset of 2.6\%
- see text).  Open circles represent observations with the 4845\AA~ $IHW$ filter extracted from the \citet{Kiselev2005} database with errors less than 1\% for various comets.  }
\label{neat_blue}
\end{figure*}

\begin{figure*}[ht]
\centering{\includegraphics[width=\textwidth]{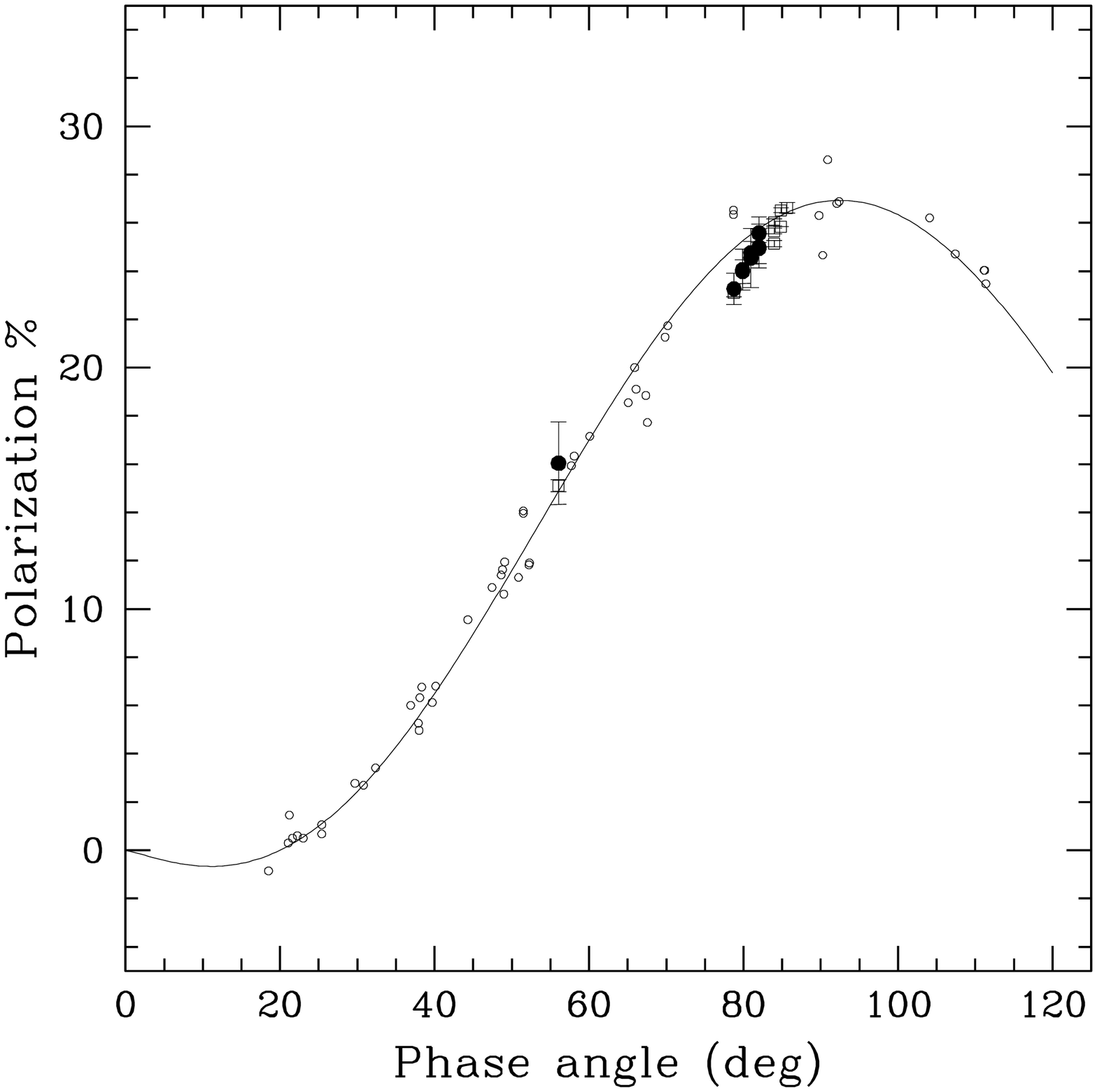}}
\caption {The phase dependence of polarization of Comet NEAT C/2001 Q4
for red wave bands centered at 6840\AA~ are shown as black solid dots with error bars.  Observations in $R$ broad band are shown as open squares (with an additional offset of 2.6\% - see text).  Open circles represent observations
with the 6840\AA~ $IHW$ filter extracted from the \citet{Kiselev2005} database with errors less than 1\% for various comets.}
\label{neat_red}
\end{figure*}

As mentioned earlier, apparition of Comet NEAT Q4, provided us the opportunity to measure
linear polarization covering a good part of the higher phase angle branch of the curve.
These measured values are plotted in the polarization phase curves for the continuum  bands
4845\AA~ and 6840\AA~ in Figs. \ref{neat_blue} and \ref{neat_red}, respectively  along with
observed values for other comets taken from the catalog compiled by \citet{Kiselev2005}.
Our observed values are in fairly good agreement with the polarization values obtained for
various comets. A large scatter is evident from the figures, more so in blue continuum
towards the larger phase angles.
On some nights observations were made through broad band $BVR$ filters (mostly with $R$-band)
to get better S/N as explained in Section 2. On May 20 and June 10, 2004,
observations have been made with both broad and narrow band filters to look for any
systematic difference in the polarization behaviour. We note such a difference in the degree of
polarization obtained in two sets of filters which has been applied to 
the broad band data ($V$ and $R$) as correction before plotting
in  Fig. \ref{neat_blue} and Fig. \ref{neat_red}, respectively. 
With this correction, the observations through $R$-band and the $IHW$ 6840\AA~band compare
well on different nights. Same is true for  the $V$ and $IHW$ 4845\AA~band apart from a
slightly larger scatter  which may be attributed to the effect of variable molecular emission
observed by \citet{Singh2006} almost simultaneously.

 The present observations indicate expected behaviour of polarization as a function of phase
angle, polarization increasing with phase in the larger phase angle regime. However, phase
angles reported here fall short of the phase values for maximal polarization.
For comet C/1996 B2 Hyakutake, \citet{Kiselev1998} report  24\% (blue continuum) and 26\% (red
continuum) as maximum polarization at $\alpha ~\sim$ 94 deg, while \citet{Manset2000} mention
a value of 28.6\% as maximum polarization in red at $\alpha ~\sim$ 91 deg for the same
comet. Though our observations do not cover the phase angles of maximal polarization,
our projected values ( $P\sim~27\% $ in the range $\alpha \sim 90-95^\circ$ ) 
for maximum P are almost similar. 

\subsection{On the nature of the grains}

We see that the polarization behaviour of comet NEAT C/2001 Q4 (Figs. \ref{neat_blue} and
\ref{neat_red}) 
is very similar to other high polarization comets, such as comet Hyakutake(C/1996 B2), at 
similar phase angles indicating that their dust grains have similar characteristics.

The grains in comets are identified to be mainly amorphous and crystalline
silicates \citep{Brownlee2006,Harker2005,Wooden2004} and CHON \citep{Kissel1986}
particles.
In comet NEAT C/2001 Q4,  Mg-rich olivine crystals
and Mg-rich crystalline pyroxene were detected in IR spectra \citep{Wooden2004}.
The mid-IR spectra of comet C/2001 Q4 (NEAT) and Hale-Bopp
\citep{Harker2002,Wooden2000,Wooden1999} show
remarkable similarity suggesting the presence of similar silicate minerals in 
approximately equal relative abundance of sub-micron grains\citep{Wooden2004}.
Thermal emission models for the 7.7 - 13.1 $\mu$m SED of comet C/2001 Q4 on two epochs
May 11.25 UT and May 11.30 UT, 2004 (just two days prior to our polarimetric
observing run) indicate the dust grains to be highly porous with size
distribution peaking at the grain radius of $a_p$ = $0.3$ $\mu$m \citep{Wooden2004}.
\citet{Wooden2004} have also reported a drop in silicate-to-amorphous
carbon by a factor of $\sim$2 in 2 hours. As far as the dust grain characteristics are
concerned, polarization observations corroborate with the finding made with
the study of thermal emission spectra.
Based on the polarization behaviour, the mean grain size of monomer is expected to 
be in the range 0.2 - 0.3 $\mu$m.

As discussed in section 3.1, the polarization color shows significant change
during the observing run indicating change in the average grain and 
aggregate characteristics.
\citet{Kolokolova2001} mention three possibilities for the change in the 
dust characteristics: i) grain size decreases due to grain fragmentation leading to 
higher abundance of smaller grains; 
ii) change in chemical compositions as volatile materials evaporate or are destroyed; 
iii) porosity increases as embedded volatile evaporate.

In the present study,
polarization color is always red, though it shows significant variation. 
One interesting finding is that the change in  polarization color is larger in smaller 
aperture. In fact the polarization color observed in the two apertures compare 
well on May 17 and May 19, but on May 18 it shows significantly lower polarization 
color and photometric color in the smaller aperture. This indicates that the comet 
has been active during  this period.

The detection of the change in polarization color  may be attributed to the change
in the composition of the dust grains rather then the change in particle size.

\section{Conclusions}
This work reported optical linear polarization observations on
comet NEAT C/2001 Q4  in the phase angle  range 86 - 55$^\circ$ where there are very few
observations. In general, polarization and brightness of the coma increase with
wavelength, phase angle and size of the apertures used.
Comet NEAT C/2001 Q4, at these large phase angles, shows similar degree of polarizations as
comet Hyakutake, indicating similar grain characteristics in two comets.  
Observations reveal that Comet NEAT Q4 exhibits polarimetric behaviour typical to high
polarization comets and therefore, should contain substantial number of small dust
particles and fluffy aggregates with submicron sized monomers.
As expected, polarimetric colour increases with wavelength and phase angle range of
present observations.  The polarization and photometric colours on May 18 indicate
some cometary activity, leading to either change in composition due to decomposition 
of organics or change in grain size. 

\section{Acknowledgements}
We thank the referees for their constructive comments and suggestions  that helped 
to improve the quality of this contribution.
The work reported here is supported by the Department of Space, Government of India.


\begin{thebibliography}{52}
\expandafter\ifx\csname natexlab\endcsname\relax\def\natexlab#1{#1}\fi
\expandafter\ifx\csname url\endcsname\relax
  \def\url#1{\texttt{#1}}\fi
\expandafter\ifx\csname urlprefix\endcsname\relax\def\urlprefix{URL }\fi

\bibitem[{{A'Hearn}(2006)}]{Ahearn2006}
{A'Hearn}, M.~F., 2006. {Whence Comets?} Science 314, 1708--1709.

\bibitem[{{Bastien} et~al.(1986){Bastien}, {Menard}, and
  {Nadeau}}]{Bastien1986}
{Bastien}, P., {Menard}, F., {Nadeau}, R., 1986. {Linear polarization
  observations of P/Halley}. \mnras ~223, 827--834.

\bibitem[{{Brooke} et~al.(1987){Brooke}, {Knacke}, and {Joyce}}]{Brooke1987}
{Brooke}, T.~Y., {Knacke}, R.~F., {Joyce}, R.~R.,  1987. {The Near Infrared
  Polarization and Color of Comet p/ Halley}. \aap ~187, 621--624.

\bibitem[{{Brownlee} et~al.(2006){Brownlee}, and {182 colleagues}}]{Brownlee2006}
{Brownlee}, D., and 182 colleagues, 2006. {Comet 81P/Wild 2 Under a Microscope}. Science
  314, 1711--1716.

\bibitem[{{Chernova} et~al.(1993){Chernova}, {Kiselev}, and
  {Jockers}}]{Chernova1993}
{Chernova}, G.~P., {Kiselev}, N.~N., {Jockers}, K.,  1993. {Polarimetric
  characteristics of dust particles as observed in 13 comets - Comparisons with
  asteroids}. Icarus 103, 144--158.

\bibitem[{{Deshpande} et~al.(1985){Deshpande}, {Joshi}, {Kulshrestha},
  {Banshidhar}, and {Vadher}}]{Deshpande1985}
{Deshpande}, M.~R., {Joshi}, U.~C., {Kulshrestha}, A.~K., {Banshidhar},
  {Vadher}, N.~M.,  1985. {An astronomical polarimeter}. Bulletin of the
  Astronomical Society of India 13, 157--161.

\bibitem[{{Dollfus}(1989)}]{Dollfus1989}
{Dollfus}, A.,  1989. {Polarimetry of grains in the coma of P/Halley. II -
  Interpretation}. \aap ~213, 469--478.

\bibitem[{{Dollfus} et~al.(1988){Dollfus}, {Bastien}, {Le Borgne},
  {Levasseur-Regourd}, and {Mukai}}]{Dollfus1988}
{Dollfus}, A., {Bastien}, P., {Le Borgne}, J.-F., {Levasseur-Regourd}, A.~C.,
  {Mukai}, T.,  1988. {Optical polarimetry of P/Halley - Synthesis of the
  measurements in the continuum}. \aap ~206, 348--356.

\bibitem[{{Dollfus} and {Suchail}(1987)}]{Dollfus1987}
{Dollfus}, A., {Suchail}, J.-L.,  1987. {Polarimetry of grains in the coma
  of P/Halley. I - Observations}. \aap ~187, 669--688.

\bibitem[{{Fulle} et~al.(2000){Fulle}, {Levasseur-Regourd}, {McBride}, and
  {Hadamcik}}]{Fulle2000}
{Fulle}, M., {Levasseur-Regourd}, A.~C., {McBride}, N., {Hadamcik}, E., 
  2000. {In Situ Dust Measurements From within the Coma of 1P/Halley:
  First-Order Approximation with a Dust Dynamical Model}. \aj ~119, 1968--1977.

\bibitem[{{Ganesh} et~al.(1998){Ganesh}, {Joshi}, {Baliyan}, and
  {Deshpande}}]{Ganesh1998}
{Ganesh}, S., {Joshi}, U.~C., {Baliyan}, K.~S., {Deshpande}, M.~R.,  1998.
  {Polarimetric observations of the comet Hale-Bopp}. \aaps ~129, 489--493.

\bibitem[{{Hadamcik} and {Levasseur-Regourd}(2003)}]{Hadamcik2003}
{Hadamcik}, E., {Levasseur-Regourd}, A.~C.,  2003. {Dust evolution of comet
  C/1995 O1 (Hale-Bopp) by imaging polarimetric observations}. \aap ~403,
  757--768.

\bibitem[{{Hadamcik} et~al.(1997){Hadamcik}, {Levasseur-Regourd}, and
  {Renard}}]{Hadamcik1997}
{Hadamcik}, E., {Levasseur-Regourd}, A.~C., {Renard}, J.~B.,  1997. {Ccd
  Polarimetric Imaging Of Comet Hale-Bopp (C/1995 O1)}. Earth Moon and Planets
  78, 365--371.

\bibitem[{{Harker} et~al.(2002){Harker}, {Wooden}, {Woodward}, and
  {Lisse}}]{Harker2002}
{Harker}, D.~E., {Wooden}, D.~H., {Woodward}, C.~E., {Lisse}, C.~M., . 2002.
  {Grain Properties of Comet C/1995 O1 (Hale-Bopp)}. \apj ~580, 579--597.

\bibitem[{{Harker} et~al.(2005){Harker}, {Woodward}, and {Wooden}}]{Harker2005}
{Harker}, D.~E., {Woodward}, C.~E., {Wooden}, D.~H.,  2005. {The Dust
  Grains from 9P/Tempel 1 Before and After the Encounter with Deep Impact}.
  Science 310, 278--280.

\bibitem[{{Jewitt}(2004)}]{Jewitt2004}
{Jewitt}, D.,  2004. {Looking through the HIPPO: Nucleus and Dust in Comet
  2P/Encke}. \aj ~128, 3061--3069.

\bibitem[{{Jockers} et~al.(2005){Jockers}, {Kiselev}, {Bonev}, {Rosenbush},
  {Shakhovskoy}, {Kolesnikov}, {Efimov}, {Shakhovskoy}, and
  {Antonyuk}}]{Jockers2005}
{Jockers}, K., {Kiselev}, N., {Bonev}, T., {Rosenbush}, V., {Shakhovskoy}, N.,
  {Kolesnikov}, S., {Efimov}, Y., {Shakhovskoy}, D., {Antonyuk}, K.,  2005.
  {CCD imaging and aperture polarimetry of comet 2P/Encke: are there two
  polarimetric classes of comets?} \aap ~441, 773--782.

\bibitem[{{Jockers} et~al.(1997){Jockers}, {Rosenbush}, {Bonev}, and
  {Credner}}]{Jockersetal1997}
{Jockers}, K., {Rosenbush}, V.~K., {Bonev}, T., {Credner}, T.,  1997.
  {Images Of Polarization And Colour In The Inner Coma Of Comet Hale-Bopp}.
  Earth Moon and Planets 78, 373--379.

\bibitem[{{Joshi} et~al.(2003){Joshi}, {Baliyan}, and {Ganesh}}]{Joshi2003}
{Joshi}, U.~C., {Baliyan}, K.~S., {Ganesh}, S.,  2003. {Polarization
  studies of comet C/2000 WM1 (LINEAR)}. \aap ~405, 1129--1135.

\bibitem[{{Joshi} et~al.(1997){Joshi}, {Baliyan}, {Ganesh}, {Chitre}, {Vats},
  and {Deshpande}}]{Joshi1997}
{Joshi}, U.~C., {Baliyan}, K.~S., {Ganesh}, S., {Chitre}, A., {Vats}, H.~O.,
  {Deshpande}, M.~R.,  1997. {A polarization study of comet Hyakutake
  (C/1996 B2).} \aap ~319, 694--698.

\bibitem[{{Joshi} et~al.(1987){Joshi}, {Deshpande}, {Sen}, and
  {Kulshrestha}}]{Joshi1987}
{Joshi}, U.~C., {Deshpande}, M.~R., {Sen}, A.~K., {Kulshrestha}, A.,  1987.
  {Polarization investigations in four peculiar supergiants with high IR
  excess}. \aap ~181, 31--33.

\bibitem[{{Kikuchi} et~al.(1987){Kikuchi}, {Mikami}, {Mukai}, {Mukai}, and
  {Hough}}]{Kikuchi1987}
{Kikuchi}, S., {Mikami}, Y., {Mukai}, T., {Mukai}, S., {Hough}, J.~H., 
  1987. {Polarimetry of Comet P/Halley}. \aap ~187, 689--692.

\bibitem[{{Kiselev} et~al.(2005){Kiselev}, {Rosenbush}, {Jockers}, {Velichko},
  and {Kikuchi}}]{Kiselev2005}
{Kiselev}, N., {Rosenbush}, V., {Jockers}, K., {Velichko}, S., {Kikuchi}, S.,
   2005. {Database of Comet Polarimetry: Analysis and Some Results}. Earth
  Moon and Planets 97, 365--378.

\bibitem[{{Kiselev} and {Velichko}(1998)}]{Kiselev1998}
{Kiselev}, N.~N., {Velichko}, F.~P.,  1998. {Polarimetry and Photometry of
  Comet C/1996 B2 Hyakutake}. Icarus 133, 286--292.

\bibitem[{{Kissel} et~al.(1986){Kissel}, and {18 colleagues}}]{Kissel1986}
{Kissel}, J., {and 18 colleagues},  1986. {Composition of comet Halley dust particles
  from Giotto observations}. \nat ~321, 336--337.

\bibitem[{{Kolokolova} et~al.(2004){Kolokolova}, {Hanner}, {Levasseur-Regourd},
  and {Gustafson}}]{Kolokolova2004}
{Kolokolova}, L., {Hanner}, M., {Levasseur-Regourd}, A.-C., {Gustafson}, P.,
  2004. {Physical properties of cometary dust from light scattering and
  emission}. In: Comets II. University of Arizona Press, pp. 577--604.

\bibitem[{{Kolokolova} et~al.(2001){Kolokolova}, {Jockers}, {Gustafson}, and
  {Lichtenberg}}]{Kolokolova2001}
{Kolokolova}, L., {Jockers}, K., {Gustafson}, B.~{\AA}.~S., {Lichtenberg}, G.,
   2001. {Color and polarization as indicators of comet dust properties and
  evolution in the near-nucleus coma}. \jgr ~106, 10113--10127.

\bibitem[{{Kolokolova} et~al.(2007){Kolokolova}, {Kimura}, {Kiselev}, and
  {Rosenbush}}]{Kolokolova2007}
{Kolokolova}, L., {Kimura}, H., {Kiselev}, N., {Rosenbush}, V.,  2007. {Two
  different evolutionary types of comets proved by polarimetric and infrared
  properties of their dust}. \aap ~463, 1189--1196.

\bibitem[{{Lamy} et~al.(1987){Lamy}, {Gruen}, and {Perrin}}]{Lamy1987}
{Lamy}, P.~L., {Gruen}, E., {Perrin}, J.~M.,  1987. {Comet P/Halley -
  Implications of the mass distribution function for the photopolarimetric
  properties of the dust coma}. \aap ~187, 767--773.

\bibitem[{{Le Borgne} et~al.(1987){Le Borgne}, {Leroy}, and
  {Arnaud}}]{Leborgne1987}
{Le Borgne}, J.~F., {Leroy}, J.~L., {Arnaud}, J.,  1987. {Polarimetry of
  Comet p/ Halley - Continuum Versus Molecular Bands}. \aap ~187, 526--530.

\bibitem[{{Levasseur-Regourd} et~al.(1986){Levasseur-Regourd}, {Bertaux},
  {Dumont}, {Festou}, {Giese}, {Giovane}, {Lamy}, {Le Blanc}, {Llebaria}, and
  {Weinberg}}]{LR1986}
{Levasseur-Regourd}, A.~C., {Bertaux}, J.~L., {Dumont}, R., {Festou}, M.,
  {Giese}, R.~H., {Giovane}, F., {Lamy}, P., {Le Blanc}, J.~M., {Llebaria}, A.,
  {Weinberg}, J.~L.,  1986. {Optical probing of comet Halley from the Giotto
  spacecraft}. \nat ~321, 341--344.

\bibitem[{{Levasseur-Regourd} et~al.(1996){Levasseur-Regourd}, {Hadamcik}, and
  {Renard}}]{LR1996}
{Levasseur-Regourd}, A.~C., {Hadamcik}, E., {Renard}, J.~B.,  1996.
  {Evidence for two classes of comets from their polarimetric properties at
  large phase angles.} \aap ~313, 327--333.

\bibitem[{{Levasseur-Regourd} et~al.(1999){Levasseur-Regourd}, {McBride},
  {Hadamcik}, and {Fulle}}]{LR1999}
{Levasseur-Regourd}, A.~C., {McBride}, N., {Hadamcik}, E., {Fulle}, M., 
  1999. {Similarities between in situ measurements of local dust light
  scattering and dust flux impact data within the coma of 1P/Halley}. \aap ~348,
  636--641.

\bibitem[{{Lisse} et~al.(2006){Lisse},  and
  {17 colleagues}}]{Lisse2006}
{Lisse}, C.~M., {and 17 colleagues},  2006. {Spitzer Spectral Observations of the Deep Impact Ejecta}.
  Science 313, 635--640.

\bibitem[{{Manset} and {Bastien}(2000)}]{Manset2000}
{Manset}, N., {Bastien}, P.,  2000. {Polarimetric Observations of Comets
  C/1995 O1 Hale-Bopp and C/1996 B2 Hyakutake}. Icarus 145, 203--219.

\bibitem[{{Mathis} et~al.(1977){Mathis}, {Rumpl}, and {Nordsieck}}]{Mathis1977}
{Mathis}, J.~S., {Rumpl}, W., {Nordsieck}, K.~H.,  1977. {The size
  distribution of interstellar grains}. \apj ~217, 425--433.

\bibitem[{{Mazets} et~al.(1986){Mazets}, and {13 colleagues}}]{Mazets1986}
{Mazets}, E.~P.,  {and 13 colleagues},  1986. {Comet Halley dust
  environment from SP-2 detector measurements}. \nat ~321, 276--278.

\bibitem[{{Metz} and {Haefner}(1987)}]{Metz1987}
{Metz}, K., {Haefner}, R.,  1987. {Circular polarization near the nucleus
  of Comet P/Halley}. \aap ~187, 539--542.

\bibitem[{{Osborn} et~al.(1990){Osborn}, {A'Hearn}, {Carsenty}, {Millis},
  {Schleicher}, {Birch}, {Moreno}, and {Gutierrez-Moreno}}]{Osborn1990}
{Osborn}, W.~H., {A'Hearn}, M.~F., {Carsenty}, U., {Millis}, R.~L.,
  {Schleicher}, D.~G., {Birch}, P.~V., {Moreno}, H., {Gutierrez-Moreno}, A.,
   1990. {Standard stars for photometry of comets}. Icarus 88, 228--245.

\bibitem[{{Petrova} et~al.(2001{\natexlab{a}}){Petrova}, {Jockers}, and
  {Kiselev}}]{Petrova2001b}
{Petrova}, E.~V., {Jockers}, K., {Kiselev}, N.~N., Sep. 2001{\natexlab{a}}. {A
  Negative Branch of Polarization for Comets and Atmosphereless Celestial
  Bodies and the Light Scattering by Aggregate Particles}. Solar System
  Research 35, 390--399.

\bibitem[{{Petrova} et~al.(2001{\natexlab{b}}){Petrova}, {Jockers}, and
  {Kiselev}}]{Petrova2001a}
{Petrova}, E.~V., {Jockers}, K., {Kiselev}, N.~N.,  2001{\natexlab{b}}.
  {Light Scattering by Aggregate Particles Comparable in Size to Wavelength:
  Application to Cometary Dust}. Solar System Research 35, 57--69.

\bibitem[{{Sandford} et~al.(2006){Sandford}, and {54 Colleagues}
  }]{Sandford2006}
{Sandford}, S.~A., {and 54 colleagues},  2006. {Organics Captured
  from Comet 81P/Wild 2 by the Stardust Spacecraft}. Science 314, 1720--1725.

\bibitem[{{Schmitz} and {Brenker}(2008)}]{Schmitz2008}
{Schmitz}, S., {Brenker}, F.~E.,  2008. {Microstructural Indications for
  Protoenstatite Precursor of Cometary MgSiO$_{3}$ Pyroxene: A Further
  High-Temperature Component of Comet Wild 2}. \apjl 681, L105--L108.

\bibitem[{{Sen} et~al.(1991){Sen}, {Joshi}, and {Deshpande}}]{Sen1991}
{Sen}, A.~K., {Joshi}, U.~C., {Deshpande}, M.~R., Dec. 1991. {Polarimetric
  properties of Comet Austin}. \mnras ~253, 738--742.

\bibitem[{{Sen} et~al.(1988){Sen}, {Joshi}, {Deshpande}, {Babu}, and
  {Kulshrestha}}]{Sen1988}
{Sen}, A.~K., {Joshi}, U.~C., {Deshpande}, M.~R., {Babu}, G.~S.~D.,
  {Kulshrestha}, A.~K.,  1988. {Polarimetric observations of comet P/Halley
  on 19 March 1986}. \aap ~204, 317--318.

\bibitem[{{Serkowski}(1974)}]{Serkowski1974}
{Serkowski}, K., 1974. {Polarimeters for Optical Astronomy}. In: {Gehrels}, T.
  (Ed.), IAU Colloq. 23: Planets, Stars, and Nebulae: Studied with
  Photopolarimetry. pp. 135.

\bibitem[{{Singh} et~al.(2006){Singh}, {Sanwal}, and {Kumar}}]{Singh2006}
{Singh}, M., {Sanwal}, B.~B., {Kumar}, B.,  2006. {Spectrophotometric study
  of the comet C/2001 Q4 (NEAT)}. Bulletin of the Astronomical Society of India
  34, 273--279.

\bibitem[{{Swamy}(1986)}]{Krisnaswamy1986}
{Swamy}, K.~S.~K., 1986. {Physics of comets}. Singapore: World Scientific
  Publication, 1986.

\bibitem[{{Wooden} et~al.(2000){Wooden}, {Butner}, {Harker}, and
  {Woodward}}]{Wooden2000}
{Wooden}, D.~H., {Butner}, H.~M., {Harker}, D.~E., {Woodward}, C.~E., 
  2000. {Mg-Rich Silicate Crystals in Comet Hale-Bopp: ISM Relics or Solar
  Nebula Condensates?} Icarus 143, 126--137.

\bibitem[{{Wooden} et~al.(1999){Wooden}, {Harker}, {Woodward}, {Butner},
  {Koike}, {Witteborn}, and {McMurtry}}]{Wooden1999}
{Wooden}, D.~H., {Harker}, D.~E., {Woodward}, C.~E., {Butner}, H.~M., {Koike},
  C., {Witteborn}, F.~C., {McMurtry}, C.~W.,  1999. {Silicate Mineralogy of
  the Dust in the Inner Coma of Comet C/1995 01 (Hale-Bopp) Pre- and
  Postperihelion}. \apj ~517, 1034--1058.

\bibitem[{{Wooden} et~al.(2004){Wooden}, {Woodward}, and {Harker}}]{Wooden2004}
{Wooden}, D.~H., {Woodward}, C.~E., {Harker}, D.~E.,  2004. {Discovery of
  Crystalline Silicates in Comet C/2001 Q4 (NEAT)}. \apjl ~612, L77--L80.

\bibitem[{{Xing} and {Hanner}(1997)}]{Xing1997}
{Xing}, Z., {Hanner}, M.~S.,  1997. {Light scattering by aggregate
  particles.} \aap ~324, 805--820.

\end{thebibliography}

\end{document}